\documentclass[pra,showpacs,twocolumn]{revtex4}%
\usepackage{amsfonts}
\usepackage{amsmath}
\usepackage{amssymb,amscd}
\usepackage{psfrag}
\usepackage{graphicx}
\usepackage{braket}
\usepackage[dvips]{epsfig}
\usepackage{epsfig}
\usepackage{subfigure}
\usepackage{color}
\usepackage{amssymb}%
\providecommand{\U}[1]{\protect\rule{.1in}{.1in}}
\begin{document}
\title{SU(1,2) interferometer}
\author{Yadong Wu}
\email{wuyadong301@sjtu.edu.cn}
\affiliation{UM-SJTU Joint Institute, Shanghai Jiao Tong University, Shanghai, 200240, PR
China }
\author{Chun-Hua Yuan}
\email{chyuan@phy.ecnu.edu.cn}
\affiliation{Quantum Institute for Light and Atoms, Department of Physics, East China
Normal University, Shanghai 200062, P. R. China}
\date{\today}

\begin{abstract}
We theoretically investigate an interferometer composed of four
four-wave-mixers by Lie group method. Lie group $SU(1,2)$ characterizes the
mode transformations of this kind of interferometer. With vacuum state
inputs, the phase sensitivity of $SU(1,2)$ interferometer achieves the
Heisenberg limit, and the absolute accuracy beats $SU(1,1)$ interferometer
because of higher intensity of light inside the interferometer. For different
input cases, the optimal combination of output photon number for detection to
obtain the best phase sensitivity is calculated. Our research on $SU(1,2)$
interferometer sheds light on the performance of $SU(1,n)$ interferometer in
quantum metrology.  

\end{abstract}

\maketitle


\section{introduction}

\label{sec:introduction} Quantum metrology
\cite{giovannetti2004quantum,giovannetti2011advances} takes the advantages of quantum
mechanics to realize the high resolution of parameter measurement in physical
systems. The original motivation of studying quantum metrology comes from
general relativity. General relativity predicts the existence of gravitational
wave, which, however, is too weak to be experimentally detected. To enhance
the sensitivity of interferometers, physicists feed high power laser into
interferometers. But it seems that the sensitivity of this kind of laser
interferometers is restricted by standard quantum limit (SQL)
\cite{PhysRevD.23.1693}. SQL indicates that the phase sensitivity is bounded
by $1/\sqrt{n}$, where $n$ is the photon number. The best phase sensitivity
which can be achieved with classical interferometer, like Mach-Zehnder
interferometer (MZI) powered by a coherent state laser, cannot go beyond the
SQL. But SQL is not the ultimate limit physicists can achieve, they develop
new methods to beat the SQL and achieve the Heisenberg limit, which indicates
that the phase sensitivity equals to $1/n$, i.e., the reciprocal of the photon
number. This kind of interferometers beating SQL are quantum interferometers.

With the development of modern mathematics, Lie group theory is widely used to
describe symmetric structures in theoretical physics.
In~\cite{PhysRevA.33.4033}, Lie group theory is used to describe the mode
transformations in an interferometer. Specifically, the interferometers like
MZI, consisting of beam splitters, can be characterized by the Lie group
$SU(2)$, while the interferometer using four wave mixers to replace beam
splitters is characterized by the Lie group $SU(1,1)$. This Lie group
structure is determined by the Lie algebra formed by the Hamiltonian of the
interferometer. With single Fock state input, the $SU(2)$ interferometer
cannot go beyond the SQL. But fed with a special correlated Fock state
$(|0,0\rangle+|0,2\rangle)/\sqrt{2}$, $SU(2)$ interferometer can attain the
Heisenberg limit~\cite{PhysRevA.33.4033}. Besides, it is shown that if the
second input of a MZI is fed with a squeezed vacuum state, the phase
sensitivity can beat the SQL~\cite{PhysRevD.23.1693}. In contrast, $SU(1,1)$
interferometer can achieve the Heisenberg limit with vacuum state inputs. It
will be much more easier to implement since it needs fewer optical devices
experimentally. But since the photon number inside the interferometer with
vacuum state inputs is relatively low, W. N. Plick et
al.~\cite{plick2010coherent} suggest injecting coherent state light into
$SU(1,1)$ interferometer to significantly enhance the phase sensitivity.
Feeding strong coherent state laser makes the photon number in $SU(1,1)$
interferometer increased vastly and makes the phase sensitivity reaching far
below the SQL. However, the phase sensitivity can only inversely scale with
the square root of the intensity of the coherent state input light. D. Li et
al.~\cite{li2014phase} proposes that injecting a coherent state and a squeezed
vacuum state into $SU(1,1)$ interferometer with homodyne detection can
decrease the noise and enhance the sensitivity. But this method cannot always
attain the Heisenberg limit, either. In this paper, we present a new
interferometer called SU(1,2) interferometer composed of four four-wave-mixers
(FWMs) which is described by the Lie group $SU(1,2)$. The absolute accuracy can
not only beat the sensitivity of $SU(1,1)$ interferometer for the same input
fields, but also make the phase sensitivity scale with the Heisenberg limit.

\begin{figure}[ptbh]
\centerline{\includegraphics[scale=0.43,angle=0]{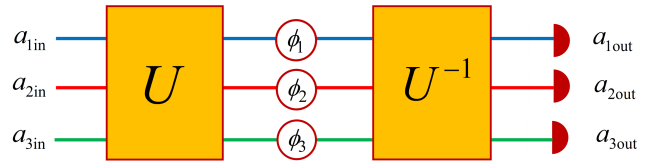}}\caption{(Color
online) General three-port balanced interferometer. $\phi_{i}$ ($i=1,2,3$):
phase shifts.}%
\label{fig1}%
\end{figure}

\begin{figure}[ptbh]
\centerline{\includegraphics[scale=0.5,angle=0]{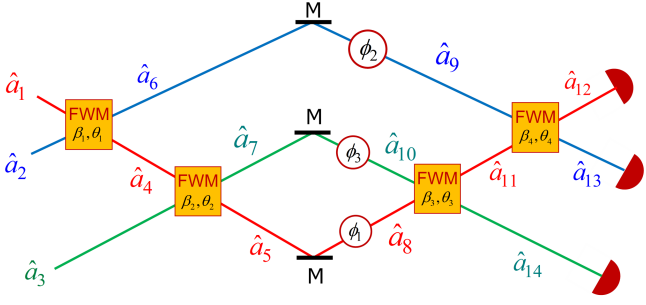}}\caption{(Color
online) The schematic diagram of the SU(1,2) interferometer. $\hat{a}_{i}$
($i=0,1,\cdots,14$) denote the different light beams in the interferometer.
$\beta_{i}$ and $\theta_{i}$ ($i=1,2,3,4$) describe the strength and phase
shift in the different four-wave-mixers (FWMs) processes. $\phi_{i}$
($i=1,2,3$): phase shifts; $M$: mirrors.}%
\label{fig2}%
\end{figure}

These interferometers we mentioned are two input and two output
interferometers, whereas three input and three output $SU(3)$ interferometer~
\cite{sanders1999vector,rowe1999representations,PhysRevLett.110.113603} and
other multi-path interferometers~
\cite{PhysRevA.46.2840,PhysRevA.55.2267,PhysRevA.55.2564,BenAryeh20102863,spagnolo2012quantum}
are also investigated for different purposes. Multi-path interferometer
generally owns more than two input and two output ports. Over two beams are
mixed in a multi-path interferometer. B. C. Sanders et
al.~\cite{sanders1999vector} investigates measuring vector phase with $SU(3)$
interferometer. $SU(3)$ interferometer, different from a two-path
interferometer, can be used to estimate two phase shifts simultaneously. G. M.
D$^{\prime}$Ariano et al.~\cite{PhysRevA.55.2267} propose a scheme of using
multi-path interferometer to enhance phase sensitivity and asserts that the
phase sensitivity can inversely scale with the number of beam paths in an
interferometer. It seems that there are two advantages of multi-path
interferometers in phase estimation. One is to estimate more phases
simultaneously, the other is to enhance the phase sensitivity. These two
advantages motivate us to investigate quantum metrology with multi-path interferometer.

As far as we know, most multi-path interferometers discussed previously uses
linear optical devices like beam splitters, while multi-path nonlinear quantum
interferometer utilizing nonlinear optical devices such as FWM has not been
theoretically investigated. We want to use the Lie group theoretical method
in~\cite{PhysRevA.33.4033} to investigate a multi-path nonlinear quantum
interferometer, denoted by $SU(1,2)$ interferometer. We're interested in the
sensitivity of $SU(1,2)$ interferometer in phase estimation and want to
compare its phase sensitivity with $SU(1,1)$
interferometer~\cite{PhysRevA.33.4033,plick2010coherent}.

This paper mainly contains three parts. Sec.~\ref{sec3} introduces eight
Hermitian operators spanning the Lie algebra $\mathfrak{su}(1,2)$. The mode
transformations given by the conjugation operation form the group $SU(1,2)$.
Then we introduce the structure of $SU(1,2)$ interferometer. Sec.~\ref{sec4}
shows that with vacuum state inputs, $SU(1,2)$ interferometer can attain the
Heisenberg limit and beats the sensitivity of $SU(1,1)$ interferometer. The
optimal combination of output photon numbers for detection is found to achieve
the best sensitivity. Sec.~\ref{sec5} analyzes the case that a coherent state
beam is fed into $SU(1,2)$ interferometer. We give the optimal combination of
output photon numbers for detection in different input cases. Furthermore, if
the coherent state light is injected into the third input, the phase
sensitivity can still approach the Heisenberg limit when the amplification
gain of the FWMs is large. Sec.~\ref{sec:con} gives the conclusion of this paper.

\section{$SU(1,2)$ Lie group and $SU(1,2)$ interferometer}

\label{sec3} In this section, based on the method of Lie group we introduce
the $SU(1,2)$ interferometer possessing three inputs and three outputs. A
general three input and three output interferometer is shown in Fig.
\ref{fig1}. If we restrict the unitary operation $U$ to $SU(1,2)$ operation,
then it is a $SU(1,2)$ interferometer. Then in Fig.~\ref{fig2} we use the
nonlinear optical devices FWMs to form one type of $SU(1,2)$ interferometer.
The connection between $SU(1,2)$ Lie group and the optical devices FWMs are
presented. It can be constructed with current experimental
technology~\cite{PhysRevLett.113.023602}.

\bigskip A $SU(1,2)$ mode transformation is
\begin{equation}
\left(
\begin{array}
[c]{c}%
\hat{a}_{1\,\mathrm{out}}\\
\hat{a}_{2\,\mathrm{out}}^{\dagger}\\
\hat{a}_{3\,\mathrm{out}}^{\dagger}%
\end{array}
\right)  =\left(
\begin{array}
[c]{ccc}%
S_{11} & S_{12} & S_{13}\\
S_{21} & S_{22} & S_{23}\\
S_{31} & S_{32} & S_{33}%
\end{array}
\right)  \left(
\begin{array}
[c]{c}%
\hat{a}_{1\,\mathrm{in}}\\
\hat{a}_{2\,\mathrm{in}}^{\dagger}\\
\hat{a}_{3\,\mathrm{in}}^{\dagger}%
\end{array}
\right)  , \label{eq0.5}%
\end{equation}
where the creative and annihilation operators satisfy the boson commutation
relations. The transformation matrix has the following relation:%
\begin{align}
&  \left(
\begin{array}
[c]{ccc}%
1 & 0 & 0\\
0 & -1 & 0\\
0 & 0 & -1
\end{array}
\right)  \left(
\begin{array}
[c]{ccc}%
S_{11} & S_{12} & S_{13}\\
S_{21} & S_{22} & S_{23}\\
S_{31} & S_{32} & S_{33}%
\end{array}
\right)  ^{\dagger}\left(
\begin{array}
[c]{ccc}%
1 & 0 & 0\\
0 & -1 & 0\\
0 & 0 & -1
\end{array}
\right) \nonumber\\
&  =\left(
\begin{array}
[c]{ccc}%
S_{11} & S_{12} & S_{13}\\
S_{21} & S_{22} & S_{23}\\
S_{31} & S_{32} & S_{33}%
\end{array}
\right)  ^{-1}. \label{eq0}%
\end{align}
All the matrices satisfying Eq.~(\ref{eq0}) form a matrix Lie group $SU(1,2)$.
To generate these $SU(1,2)$ mode transformations, we introduce eight Hermitian
operators which are given by
\begin{align}
&  K_{1}=\frac{1}{2}(\hat{a}_{1}^{\dagger}\hat{a}_{2}^{\dagger}+\hat{a}%
_{1}\hat{a}_{2}),\text{ }K_{2}=-\frac{i}{2}(\hat{a}_{1}^{\dagger}\hat{a}%
_{2}^{\dagger}-\hat{a}_{1}\hat{a}_{2}),\nonumber\\
&  K_{3}=\frac{1}{2}(\hat{a}_{1}^{\dagger}\hat{a}_{3}^{\dagger}+\hat{a}%
_{1}\hat{a}_{3}),\text{ }K_{4}=-\frac{i}{2}(\hat{a}_{1}^{\dagger}\hat{a}%
_{3}^{\dagger}-\hat{a}_{1}\hat{a}_{3}),\nonumber\\
&  K_{5}=-\frac{1}{2}(\hat{a}_{2}^{\dagger}\hat{a}_{3}+\hat{a}_{3}^{\dagger
}\hat{a}_{2}),K_{6}=-\frac{i}{2}(\hat{a}_{2}^{\dagger}\hat{a}_{3}-\hat{a}%
_{3}^{\dagger}\hat{a}_{2}),\nonumber\\
&  K_{7}=\frac{1}{2}(\hat{a}_{1}^{\dagger}\hat{a}_{1}+\hat{a}_{2}\hat{a}%
_{2}^{\dagger}),\text{ }K_{8}=\frac{1}{2\sqrt{3}}(\hat{a}_{1}^{\dagger}\hat
{a}_{1}-\hat{a}_{2}\hat{a}_{2}^{\dagger}+2\hat{a}_{3}\hat{a}_{3}^{\dagger}).
\label{eqs1}%
\end{align}
Each operator $K_{i}$ ($i=1,2,\cdots,8$) in Eqs. (\ref{eqs1}) is a Hamiltonian
of some physical process generating $SU(1,2)$ mode transformations.
$K_{1},K_{2},K_{3}$ and $K_{4}$ create or annihilate photons in pairs, which
describe the process of generating photons in FWMs. $K_{1}$ and $K_{2}$ are
the Hamiltonian of a FWM with beam $1$ and beam $2$ as inputs, and $K_{3}$ and
$K_{4}$ are the Hamiltonian of a FWM with beam $1$ and beam $3$ as inputs.
$K_{5}$ and $K_{6}$ annihilate a photon and create another photon
simultaneously, maintaining the total photon number. It is a passive optical
device, generally we call beam splitter. $K_{5}$ and $K_{6}$ are the
Hamiltonian of a beam splitter with beam $2$ and beam $3$ as inputs. $K_{7}$
and $K_{8}$ are combinations of photon number operators and numbers. They can
be the Hamiltonian of relative phase shifts.

Using Baker-Campbell-Hausdorff (BCH) formula, we can
calculate the conjugation of $\hat{a}_{1},\hat{a}_{2}^{\dagger}$ and $\hat
{a}_{3}^{\dagger}$ given by the exponential of the operators in Eqs.
(\ref{eqs1}), which are given by
\begin{align}
e^{i\alpha_{1}K_{1}}M_{A}e^{-i\alpha_{1}K_{1}}  &  =\left(
\begin{array}
[c]{ccc}%
\cosh{\frac{\alpha_{1}}{2}} & -i\sinh{\frac{\alpha_{1}}{2}} & 0\\
i\sinh{\frac{\alpha_{1}}{2}} & \cosh{\frac{\alpha_{1}}{2}} & 0\\
0 & 0 & 1
\end{array}
\right)  M_{A},\nonumber\\
e^{i\alpha_{2}K_{2}}M_{A}e^{-i\alpha_{2}K_{2}}  &  =\left(
\begin{array}
[c]{ccc}%
\cosh{\frac{\alpha_{2}}{2}} & -\sinh{\frac{\alpha_{2}}{2}} & 0\\
-\sinh{\frac{\alpha_{2}}{2}} & \cosh{\frac{\alpha_{2}}{2}} & 0\\
0 & 0 & 1
\end{array}
\right)  M_{A},\nonumber
\end{align}%
\begin{align}
e^{i\alpha_{3}K_{3}}M_{A}e^{-i\alpha_{3}K_{3}}  &  =\left(
\begin{array}
[c]{ccc}%
\cosh{\frac{\alpha_{3}}{2}} & 0 & -i\sinh{\frac{\alpha_{3}}{2}}\\
0 & 1 & 0\\
i\sinh{\frac{\alpha_{3}}{2}} & 0 & \cosh{\frac{\alpha_{3}}{2}}%
\end{array}
\right)  M_{A},\nonumber\\
e^{i\alpha_{4}K_{4}}M_{A}e^{-i\alpha_{4}K_{4}}  &  =\left(
\begin{array}
[c]{ccc}%
\cosh{\frac{\alpha_{4}}{2}} & 0 & -\sinh{\frac{\alpha_{4}}{2}}\\
0 & 1 & 0\\
-\sinh{\frac{\alpha_{4}}{2}} & 0 & \cosh{\frac{\alpha_{4}}{2}}%
\end{array}
\right)  M_{A},\nonumber
\end{align}%
\begin{align}
e^{i\alpha_{5}K_{5}}M_{A}e^{-i\alpha_{5}K_{5}}  &  =\left(
\begin{array}
[c]{ccc}%
1 & 0 & 0\\
0 & \cos{\frac{\alpha_{5}}{2}} & -i\sinh{\frac{\alpha_{5}}{2}}\\
0 & -i\sin{\frac{\alpha_{5}}{2}} & \cos{\frac{\alpha_{5}}{2}}%
\end{array}
\right)  M_{A},\nonumber\\
e^{i\alpha_{6}K_{6}}M_{A}e^{-i\alpha_{6}K_{6}}  &  =\left(
\begin{array}
[c]{ccc}%
1 & 0 & 0\\
0 & \cos{\frac{\alpha_{6}}{2}} & -\sinh{\frac{\alpha_{6}}{2}}\\
0 & \sin{\frac{\alpha_{6}}{2}} & \cos{\frac{\alpha_{6}}{2}}%
\end{array}
\right)  M_{A},\nonumber
\end{align}%
\begin{align}
e^{i\alpha_{7}K_{7}}M_{A}e^{-i\alpha_{7}K_{7}}  &  =\left(
\begin{array}
[c]{ccc}%
e^{-i\frac{\alpha_{7}}{2}} & 0 & 0\\
0 & e^{i\frac{\alpha_{7}}{2}} & 0\\
0 & 0 & 1
\end{array}
\right)  M_{A},\nonumber\\
e^{i\alpha_{8}K_{8}}M_{A}e^{-i\alpha_{8}K_{8}}  &  =\left(
\begin{array}
[c]{ccc}%
e^{-i\frac{\alpha_{8}}{2\sqrt{3}}} & 0 & 0\\
0 & e^{-i\frac{\alpha_{8}}{2\sqrt{3}}} & 0\\
0 & 0 & e^{i\frac{\alpha_{8}}{\sqrt{3}}}%
\end{array}
\right)  M_{A}, \label{eq2}%
\end{align}
where $M_{A}=\left(
\begin{array}
[c]{c}%
\hat{a}_{1}\\
\hat{a}_{2}^{\dagger}\\
\hat{a}_{3}^{\dagger}%
\end{array}
\right)  $. The transformations are in the form of Eq. (\ref{eq0.5}), and
satisfy Eq. (\ref{eq0}), implying that the transformations are all $SU(1,2)$
matrices. From Eqs. (\ref{eq2}), we can see that $e^{i\alpha_{1}K_{1}}$ and
$e^{i\alpha_{2}K_{2}}$ amplify beam $1$ and beam $2$ and that $e^{i\alpha
_{3}K_{3}}$ and $e^{i\alpha_{4}K_{4}}$ amplify beam $1$ and beam $3$. They
represent the operations of FWMs. $e^{i\alpha_{5}K_{5}}$ and $e^{i\alpha
_{6}K_{6}}$ generate a combination of beam $2$ and beam $3$, which is like a
rotation in the two-dimensional space spanned by $\hat{a}_{2}$ and $\hat
{a}_{3}$. They represent the operations of beam splitters. Whereas
$e^{i\alpha_{7}K_{7}}$ and $e^{i\alpha_{8}K_{8}}$ multiply $\hat{a}_{1}%
,\hat{a}_{2}$ and $\hat{a}_{3}$ by a unit complex number. This process only
changes the phase of each mode.

The eight operators in Eqs. (\ref{eqs1}) span the Lie algebra $\mathfrak{su}%
(1,2)$. From the Lie group theory~\cite{cahn2006semi} we know that the
conjugation given by the exponential of a Lie algebra forms a Lie group.
Specifically, in our case, $K_{i}$ ($i$ is from $1$ to $8$) form the Lie
algebra $\mathfrak{su}(1,2)$, the conjugation given by the exponential of
$K_{i}$ in the vector space spanned by $\hat{a}_{1}, \hat{a}_{2}^{\dagger}$
and $\hat{a}_{3}^{\dagger}$ form the Lie group $SU(1,2)$.

It's interesting that the combination of photon number operators $\hat{a}%
_{1}^{\dagger}\hat{a}_{1}-\hat{a}_{2}^{\dagger}\hat{a}_{2}-\hat{a}%
_{3}^{\dagger}\hat{a}_{3}$ commute with all the $K_{i}$. It means that
throughout any $SU(1,2)$ mode transformation, the photon number difference
$\hat{n}_{1}-\hat{n}_{2}-\hat{n}_{3}$ remains the same. It makes sense because
the FWMs generate photon pairs in beam $1$ and beam $2$, or in beam $1$ and
beam $3$, while beam splitter remain the total photon number in beam $2$ and
beam $3$ invariant. Going through any combination of these devices, the same
amount of photons will be added into beam $1$ and the combination of beam $2$
and beam $3$.

A $SU(1,2)$ interferometer consisting of four FWMs is shown in Fig.
\ref{fig2}. The two inputs of a FWM are different: one is called probe beam,
the other one is idler beam. We use beam $1$ and beam $2$ to denote the probe
beam and the idler beam of the first FWM respectively. The two output beams
are denoted by beam $4$ and beam $6$ :
\begin{align*}
\hat{a}_{4}  &  =\cosh\frac{\beta_{1}}{2}\hat{a}_{1}+e^{-i\theta_{1}}%
\sinh\frac{\beta_{1}}{2}\hat{a}_{2}^{\dagger},\qquad\\
\hat{a}_{6}  &  =\cosh\frac{\beta_{1}}{2}\hat{a}_{2}+e^{-i\theta_{1}}%
\sinh\frac{\beta_{1}}{2}\hat{a}_{1}^{\dagger},
\end{align*}
where $\beta_{1}$ and $\theta_{1}$ are the amplification gain and the phase
parameter of the first FWM. Later we use $\theta_{j}$ and $\beta_{j}$ ($j$ is
from $1$ to $4$) to denote the phase parameters and the amplification
parameters of each FWM from left to right. Then beam $4$ is injected into the
second FWM as the probe beam, and beam $3$ injected as the idler beam. The two
output beams are denoted by beam $5$ and beam $7$:
\begin{align*}
\hat{a}_{5}  &  =\cosh\frac{\beta_{2}}{2}\hat{a}_{4}+e^{-i\theta_{2}}%
\sinh\frac{\beta_{2}}{2}\hat{a}_{3}^{\dagger},\qquad\\
\hat{a}_{7}  &  =\cosh\frac{\beta_{2}}{2}\hat{a}_{3}+e^{-i\theta_{2}}%
\sinh\frac{\beta_{2}}{2}\hat{a}_{4}^{\dagger}.
\end{align*}
After that, beam $5$, $6$ and $7$ experience three independent phase shifts,
$\phi_{1}$, $\phi_{2}$ and $\phi_{3}$, respectively. We denote the three beams
after phase shifting as beam $8$ to beam $10$. Then they pass through the
third and the fourth FWMs. The transformations of the annihilation operators
in each beam can be written as
\begin{align}
\hat{a}_{12}  &  =\cosh\frac{\beta_{4}}{2}\hat{a}_{11}+e^{-i\theta_{4}}%
\sinh\frac{\beta_{4}}{2}\hat{a}_{9}^{\dagger},\nonumber\\
\hat{a}_{13}  &  =\cosh\frac{\beta_{4}}{2}\hat{a}_{9}+e^{-i\theta_{4}}%
\sinh\frac{\beta_{4}}{2}\hat{a}_{11}^{\dagger},\qquad\nonumber\\
\qquad\hat{a}_{11}  &  =\cosh\frac{\beta_{3}}{2}\hat{a}_{8}+e^{-i\theta_{3}%
}\sinh\frac{\beta_{3}}{2}\hat{a}_{10}^{\dagger},\nonumber\\
\hat{a}_{14}  &  =\cosh\frac{\beta_{3}}{2}\hat{a}_{10}+e^{-i\theta_{3}}%
\sinh\frac{\beta_{3}}{2}\hat{a}_{8}^{\dagger},.
\end{align}
In the end, one can detect the sum of the output photon numbers in beam $12$
and $13$\ and the photon number in beam $14$.

In the following, we are going to investigate the phase sensitivity of
$SU(1,2)$ interferometer. The phase sensitivity is defined by
\begin{equation}\label{eq6}
(\Delta\phi_{j})_{sn_{12}+tn_{13}+rn_{14}}=\frac{\Delta(s\hat{n}_{12}+t\hat
{n}_{13}+r\hat{n}_{14})}{\left\vert \frac{\partial\langle s\hat{n}_{12}%
+t\hat{n}_{13}+r\hat{n}_{14}\rangle}{\partial\phi_{j}}\right\vert },
\end{equation}
where $j$ can be $1,2,$ or $3$, $s,t,r$ are real numbers, and $\hat{n}_{12}%
=\hat{a}_{12}^{\dagger}\hat{a}_{12}$, similar for $\hat{n}_{13}$ and $\hat
{n}_{14}$, denoting the photon
numbers in beam $12,\,13,\,14$. Phase sensitivity is the standard deviation of
the estimation result of the phase shift. Choose the output photon number as
the estimator. The phase sensitivity equals to the standard deviation of the
estimator divided by the derivative of the mean value of the estimator with
respect to the corresponding phase shift. Various estimators result in
different phase sensitivities for the same estimation target.

Without loss of generality, we set the phase parameters of the FWMs
$\theta_{1}=\theta_{2}=0$ and $\theta_{3}=\theta_{4}=\pi$. The $\pi$
change of the phase parameters makes the third and fourth FWMs perform the
inverse operations of the first and second FWMs, respectively.

\section{Phase sensitivity with vacuum state inputs}

\label{sec4} This section investigates the phase sensitivity of $SU(1,2)$
interferometer when all the three inputs are vacuum states.

The detailed calculation of the phase sensitivity is omitted. Here we only focus on analysis of the results. We find that
the optimal sensitivity is achieved when the sum of the photon numbers in beam
$12$ and beam $14$ is detected. The comparison of the phase sensitivity of
$SU(1,2)$ interferometer with $SU(1,1)$ interferometer shows its advantage of
enhancing phase sensitivity.

\begin{figure}[tbh]
\includegraphics[width=0.45\textwidth]{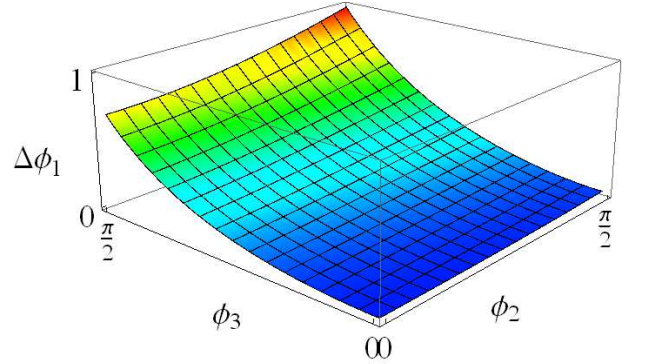} \caption{(Color online) Phase
sensitivity $\Delta\phi_{1}$ is shown as a function of phase shifts $\phi_{2}$
and $\phi_{3}$, when $\beta_1=\beta_2=3$.}%
\label{fig3}%
\end{figure}

In $SU(1,2)$ interferometer, each phase sensitivity depends on the amount of
the phase shift in each beam. Each phase sensitivity achieves its minimum when
all the three phase shifts vanish, when the output state approaches the input
state. Fig. \ref{fig3} shows how the phase sensitivity $\Delta\phi_{1}$ varies
with respect to the other two phase shifts. The impact of shifting $\phi_{3}$
on the sensitivity $\Delta\phi_{1}$ is much greater than shifting $\phi_{2}$.
Because
\begin{align}
\hat{a}_{5}  &  =\cosh\frac{\beta_{1}}{2}\cosh\frac{\beta_{2}}{2}\hat{a}%
_{1}+\sinh\frac{\beta_{1}}{2}\cosh\frac{\beta_{2}}{2}\hat{a}_{2}^{\dagger
}+\sinh\frac{\beta_{2}}{2}\hat{a}_{3}^{\dagger},\\
\hat{a}_{7}  &  =\cosh\frac{\beta_{1}}{2}\sinh\frac{\beta_{2}}{2}\hat{a}%
_{1}^{\dagger}+\sinh\frac{\beta_{1}}{2}\sinh\frac{\beta_{2}}{2}\hat{a}%
_{2}+\cosh\frac{\beta_{2}}{2}\hat{a}_{3}.
\end{align}
For large $\beta_{2}$, $\hat{a}_{5}$ is nearly equal to $\hat{a}_{7}^{\dagger
}$. Thus shifting $\phi_{3}$ with $\delta\phi$ is nearly equivalent to
shifting $\phi_{1}$ with $-\delta\phi$, which significantly affect $\Delta
\phi_{1}$.

Since the phase shift can be controlled by feedback
loops~\cite{yonezawa2012quantum} to keep the phase sensitivity optimal, we
assume during each single phase estimation, the other two phase shifts vanish.
We will focus on the phase sensitivity $\Delta\phi_{1}$ or $\Delta\phi_{3}$.
As for $\Delta\phi_{2}$, when $\phi_{1}=\phi_{3}=0$, the $SU(1,2)$
interferometer is similar to the $SU(1,1)$ interferometer, but the light
intensities insider the interferometers are different.

\begin{figure}[ptbh]
\center\includegraphics[width=0.45\textwidth]{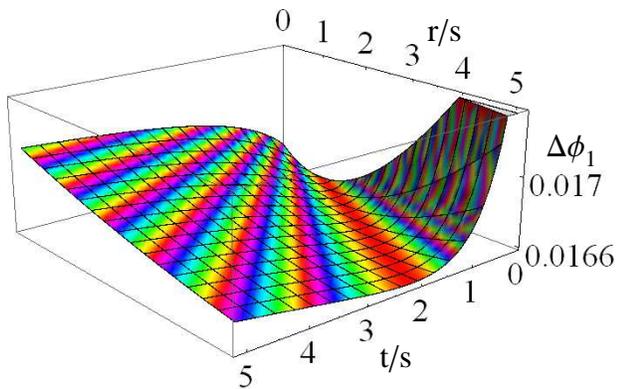} \caption{(Color online)
The phase sensitivity $(\Delta\phi_{1})_{s n_{12}+t n_{13}+r n_{14}%
}$ is plotted as a function of parameter ratios $t/s$ and
$r/s$. Set $\beta_{1}=\beta_{2}=3$.}%
\label{fig4}%
\end{figure}

At the output side, detecting different combinations of photon numbers in
three outputs may lead to different phase sensitivities, among which, one
combination is optimal. Fig. \ref{fig4} shows the phase sensitivity
$\Delta\phi_{1}$ as a function of the ratios $t/s$ and $r/s$. Any combination of the output photon
numbers can be written as a combination of the linearly independent operators
$\hat{n}_{12}+\hat{n}_{13},\hat{n}_{12}-\hat{n}_{13}+2\hat{n}_{14}$ and
$\hat{n}_{12}-\hat{n}_{13}-\hat{n}_{14}$. The first two operators can be
obtained from $K_{7}$ and $K_{8}$ at the output side, and the last one is the
invariant operator equaling to $\hat{n}_{1}-\hat{n}_{2}-\hat{n}_{3}$. For
vacuum state input, $\langle\hat{n}_{12}-\hat{n}_{13}-\hat{n}_{14}%
\rangle=Var(\hat{n}_{12}-\hat{n}_{13}-\hat{n}_{14})=0$. So we only need to
find the optimal ratio between $\hat{n}_{12}+\hat{n}_{13}$ and $\hat{n}%
_{12}-\hat{n}_{13}+2\hat{n}_{14}$. It's found that the best combination is the
sum of the output photon numbers in beam $12$ and beam $14$, i.e., $\hat
{n}_{12}+\hat{n}_{14}$, which is achieved when $r/s=1$ and $t/s=0$ in Fig.~\ref{fig4}.

By taking the limitations of the expression of phase sensitivity at zero phase
shifts, we obtain the optimal phase sensitivity $(\Delta\phi_{1}%
)_{n_{12}+n_{14}}$ which is given by
\begin{equation}
(\Delta\phi_{1})_{n_{12}+n_{14}}\sim\frac{2}{\cosh\beta_{1}\cosh\beta_{2}}.
\label{eq8}%
\end{equation}
$(\Delta\phi_{1})_{n_{12}+n_{13}}$ and $(\Delta\phi_{3})_{n_{12}+n_{13}}$ are
not exactly the same, but when $\cosh\beta_{1}, \cosh\beta_{2}\gg1$, they
are nearly equal to each other. The phase sensitivity of $SU(1,1)$
interferometer is $1/\sinh\beta$, where $\beta$ is the amplification gain of
the FWMs in $SU(1,1)$ interferometer. Because of a second optical parametric
amplification, the phase sensitivity $(\Delta\phi_{1})_{n_{12}+n_{14}}$ and
$(\Delta\phi_{3})_{n_{12}+n_{14}}$ are improved compared to the $SU(1,1)$
interferometer. With vacuum states input, the absolute accuracy of $SU(1,2)$
interferometer beats that of $SU(1,1)$ interferometer.

\begin{figure}[ptbh]
\center
\subfigure[]{
\includegraphics[width=0.4\textwidth]{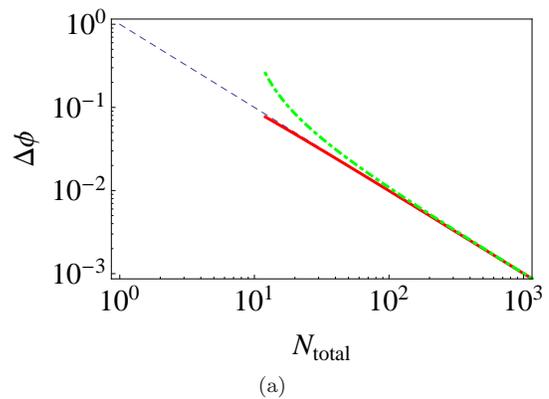}}\qquad\subfigure[]{
\includegraphics[width=0.4\textwidth]{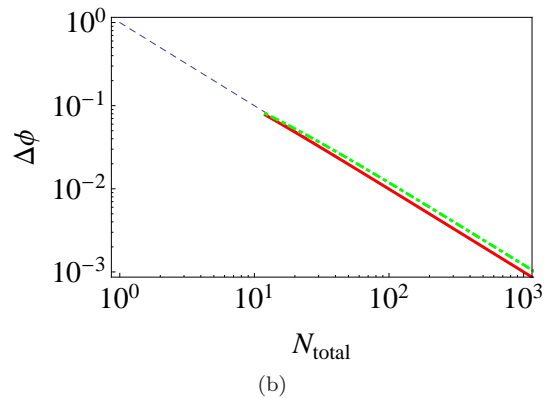}}
\caption{(Color online) The phase sensitivities $\Delta\phi$ of the $SU(1,1)$
interferometer versus the total photon number $N_{\text{total}}$ with parameter (a) $\beta_{1}=3$; (b) $\beta_{2}=3$.
$(\Delta\phi_{1})_{n_{12}+n_{14}}$ and $(\Delta\phi_{3})_{n_{12}+n_{14}}$ are described by the solid red line and the dot-dashed green curve, respectively. The dashed blue line
is the Heisenberg limit.}
\label{fig5}%
\end{figure}

In order to know whether the phase sensitivity can achieve the Heisenberg
limit, we need to find how the phase sensitivities scale with the photon
number. The total photon number in the interferometry is the sum of the photon
numbers in beam $5$, beam $6$ and beam $7$ as shown in Fig. \ref{fig2}, which
is given by%
\begin{equation}
N_{\text{total}}=\sinh^{2}\frac{\beta_{1}}{2}(\cosh^{2}\frac{\beta_{2}}%
{2}+1)+\sinh^{2}\frac{\beta_{2}}{2}(\cosh^{2}\frac{\beta_{1}}{2}%
+1).\label{eq3}%
\end{equation}
The phase sensitivities $(\Delta\phi_{1})_{n_{12}+n_{14}})$
(solid red line) and $(\Delta\phi_{3})_{n_{12}+n_{14}}$
(dot-dashed green curve) as functions of the total photon number
$N_{\text{total}}$ are shown in Fig. \ref{fig5}. The dashed blue line
represents the Heisenberg limit. It can be seen that both $(\Delta\phi
_{1})_{n_{12}+n_{14}}$ and $(\Delta\phi_{3})_{n_{12}+n_{14}%
}$ nearly equal to the Heisenberg limit.

\section{Phase sensitivity with coherent state input}

\label{sec5} This section still investigates the $SU(1,2)$ interferometer in
Fig. \ref{fig2}. Whereas the inputs become one coherent state $|\alpha\rangle$
and two vacuum states. There are three different cases since a coherent state
can be injected into any one of the three input ports. We will find that with
one coherent state input, the phase sensitivity of $SU(1,2)$ interferometer
can be further enhanced and it's still possible for the phase sensitivity to
achieve the Heisenberg limit.

\begin{figure}[ptbh]
\center
\includegraphics[width=0.4\textwidth]{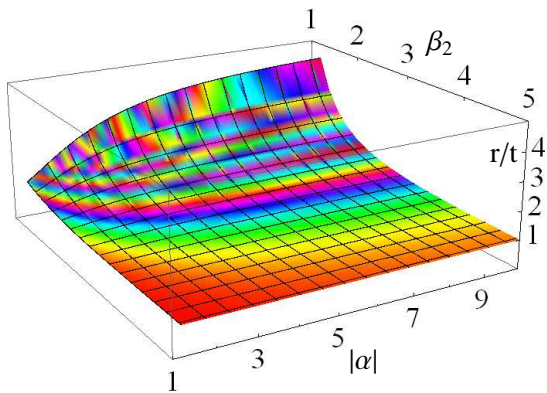}\caption{(Color online) The optimal
value of $r/t$ as a function of the amplification gain of FWMs $\beta_{2}$ and
the intensity of the coherent state input $|\alpha|$ is plotted.}%
\label{fig6}%
\end{figure}

\begin{figure}[ptbh]
\center
\includegraphics[width=0.4\textwidth]{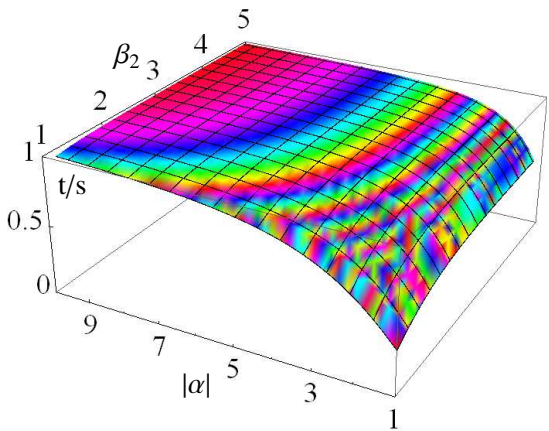}\caption{(Color online) The optimal
value of $t/s$ as a function of the amplification gain of FWMs $\beta_{2}$ and
the intensity of the coherent state input $|\alpha|$ is plotted.}%
\label{fig7}%
\end{figure}

When a coherent state beam is injected into input $1$, to achieve the optimal
phase sensitivity at zero phase shifts, one cannot choose the photon number
$n_{12}$ as the estimator to estimate any phase shift, otherwise the phase
sensitivity will diverge. So $s=0$ in Eq.~\ref{eq6} and we can only detect the combination of
photon numbers in beam $13$ and beam $14$. Which linear combination of
$n_{13}$ and $n_{14}$ leads to the optimal phase sensitivity depends on the
amplification gain of the FWMs and the intensity of the coherent state light
input. Fig. \ref{fig6} shows the optimal value of $r/t$ to obtain the best
phase sensitivity $\Delta\phi_{1}$ when detecting the combination $t n_{13}+r
n_{14}$. When the intensity of the coherent state input is low or the
amplification gain of the FWMs is high, the best combination is $n_{13}%
+n_{14}$. Otherwise, the optimal $r$ will be larger than $1$. If the coherent
state light is injected into the input $2$, then one need to detect the
combination of photon numbers in beam $12$ and beam $14$. It's similar for the
case that the coherent state is in input $3$. Fig. \ref{fig7} shows the
optimal ratio $t/s$ when a coherent state light is fed into input $3$.

\begin{figure}[ptbh]
\subfigure[]{
\includegraphics[width=0.23\textwidth]{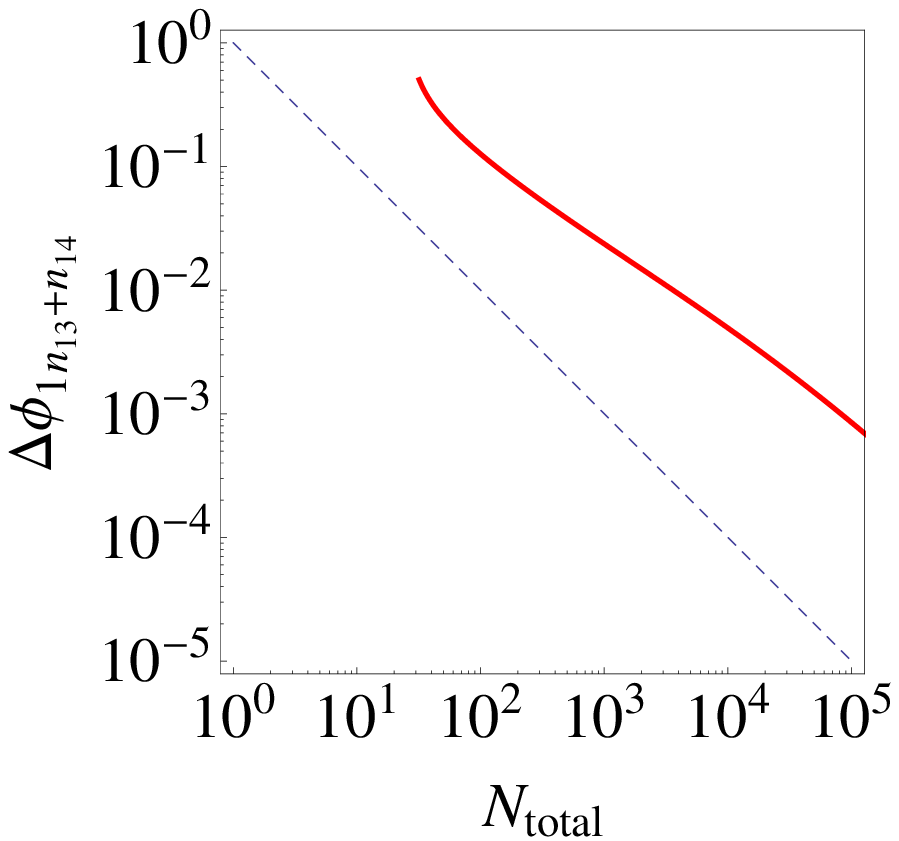}} \subfigure[]{
\includegraphics[width=0.23\textwidth]{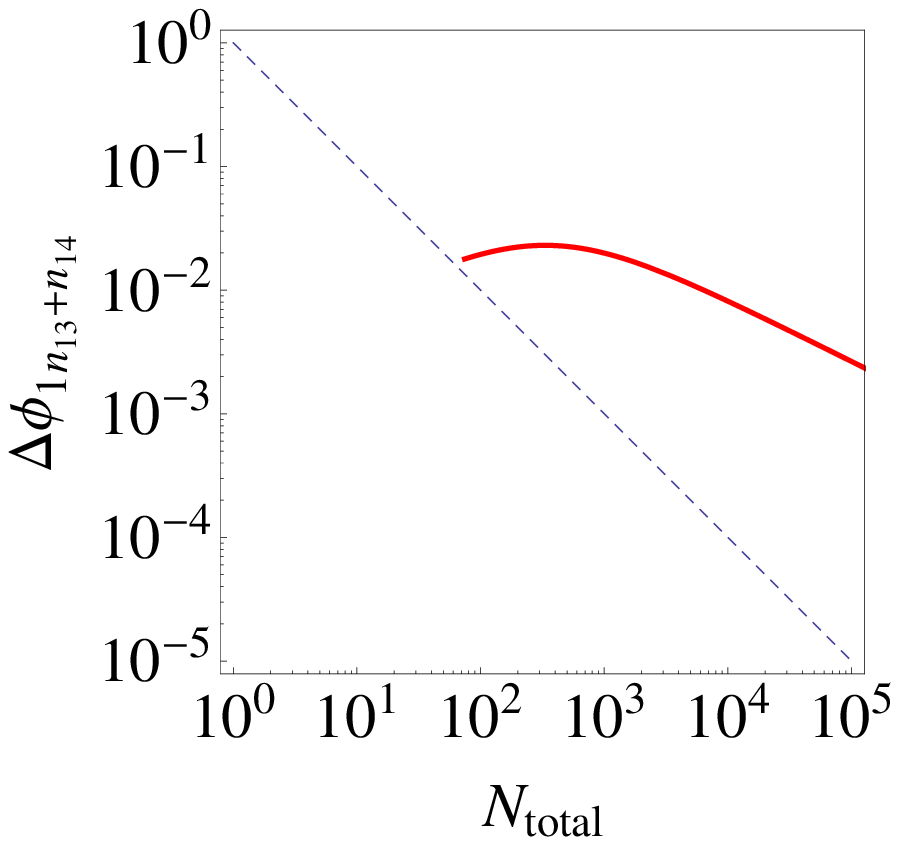}} \newline\subfigure[]{
\includegraphics[width=0.23\textwidth]{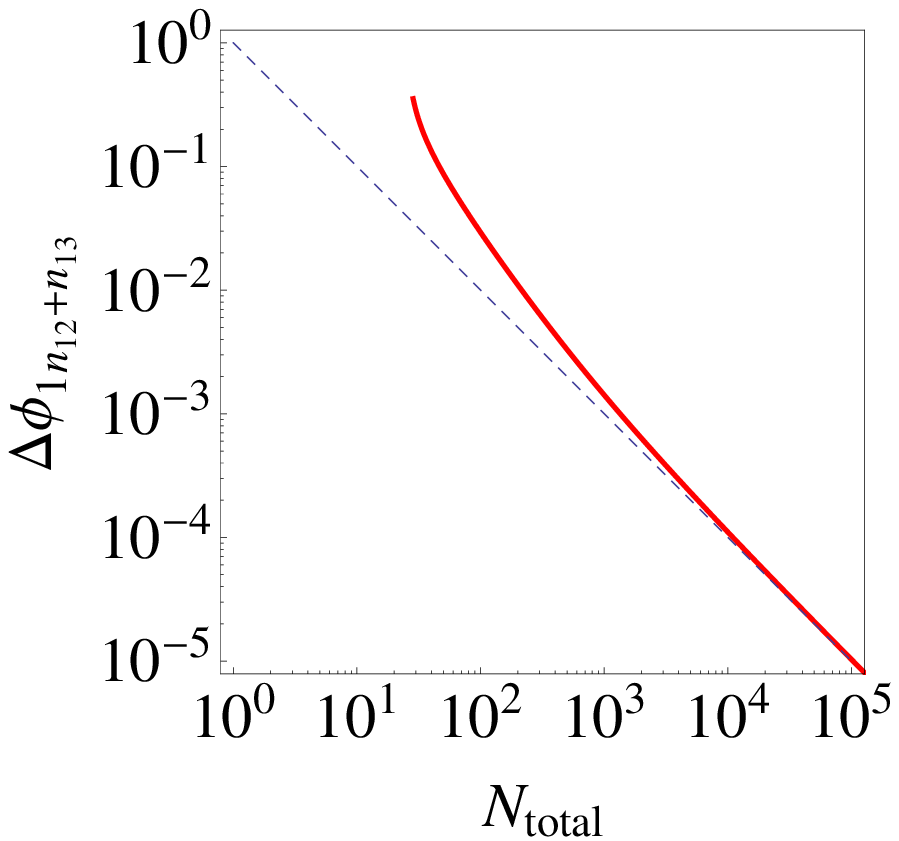}} \subfigure[]{
\includegraphics[width=0.23\textwidth]{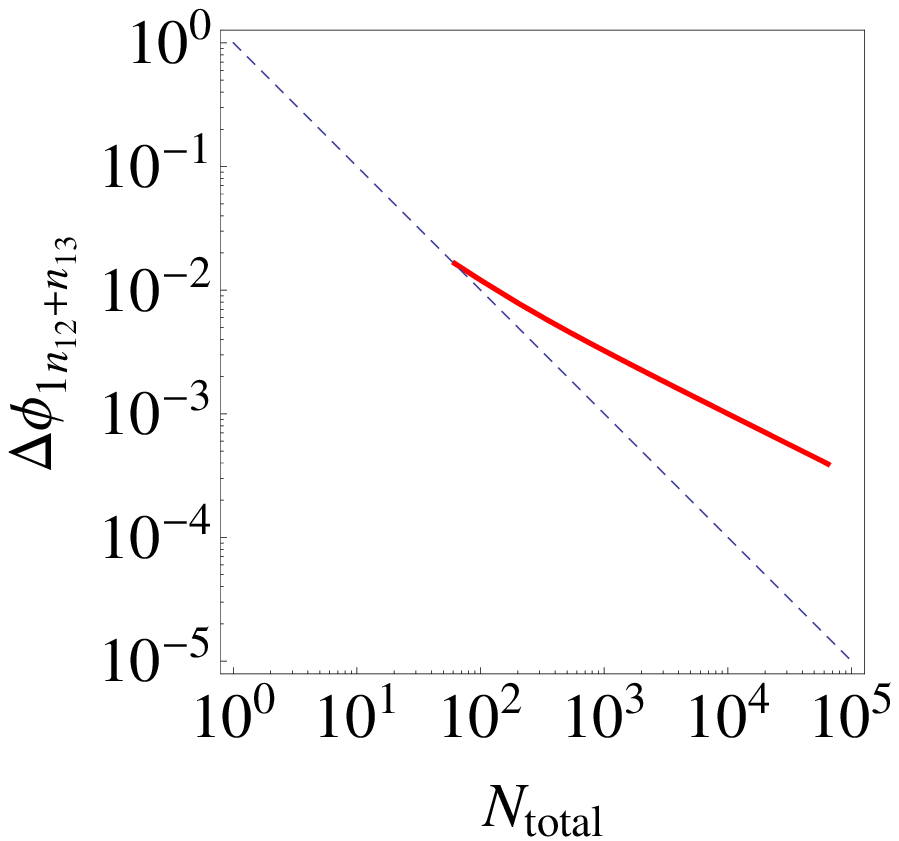}}
\caption{ The phase sensitivities $(\Delta\phi_{1})_{n_{13}+n_{14}}$ and $(\Delta\phi_{1})_{n_{12}+n_{13}}$ as function of the total number $N_{total}$ in figure (a) and (b) with the coherent light feeding into the first input, and in figure (c) and (d) with the coherent light feeding into the third input, respectively. The dashed blue lines
represent the Heisenberg limit. In (a) and (c), set $|\alpha|=5, \beta_{1}=\beta_{2}$, and vary the
values of $\beta_{1}$ and $\beta_{2}$. In (b) and (d), set $\beta_{1}%
=\beta_{2}=3$ and vary the value of $|\alpha|$.}%
\label{fig8}%
\end{figure}

Lastly, let's see whether the phase sensitivity of $SU(1,2)$ interferometer
with one coherent state input can achieve the Heisenberg limit. In the case of
one coherent state light input, the total photon number will be different.
When the coherent state light is fed into input $1$ or input $3$, the phase
sensitivity $(\Delta\phi_{1})_{n_{13}+n_{14}}$ and $(\Delta\phi_{1}%
)_{n_{12}+n_{13}}$ are respectively shown in Fig. \ref{fig8} as a function of
photon number $N_{total}$. It's shown that if the coherent state light is
injected into the first input, the phase sensitivity cannot achieve the
Heisenberg limit, while if the coherent state light is fed into the third
input, it can attain the Heisenberg limit when the amplification gains are
large. Comparing the values of the phase sensitivities in these two cases, we
find injecting coherent state light into the third input leads to better phase sensitivity.

\section{Discussion and Conclusion}

\label{sec:con}

This paper discusses the ideal model of $SU(1,2)$ interferometer, since photon
number loss error has not been considered~\cite{PhysRevA.86.023844}. But the
phenomenon of photon loss widely exists in practical FWMs. $SU(1,2)$
interferometers utilizes two more FWMs than $SU(1,1)$ interferometers, which
necessarily introduces higher probability of photon loss. In practical
experiments, whether the enhancement of phase sensitivity of $SU(1,2)$
interferometer is robust to the photon loss error has not been studied now.

In this paper, we show that $SU(1,2)$ interferometer can enhance the phase
sensitivity compared with $SU(1,1)$ interferometer. This work sheds light on
the performance of $SU(1,n)$ interferometer. We believe by adding more FWMs,
$SU(1,n)$ interferometer can further enhance the phase sensitivity. What's
more, in this work, we use the Lie group theoretical method to investigate
quantum interferometers. The way of our investigation in this paper
can be generalized into any $SU(m,n)$ interferometer.

In conclusion, this paper uses Lie group and Lie algebra theory to analyze
one multi-path nonlinear optical quantum interferometer, denoted by $SU(1,2)$
interferometer. Eight Hermitian operators spanning Lie algebra $\mathfrak{su}%
(1,2)$ are introduced to describe the Hamiltonian of $SU(1,2)$ interferometer.
We calculate and analyze the phase sensitivities in $SU(1,2)$ interferometer.
With all vacuum state inputs, the optimal phase sensitivity is achieved when
the sum of the photon numbers in beam $12$ and beam $14$ is detected. This
phase sensitivity of $SU(1,2)$ interferometer can achieve the Heisenberg limit
and beats the sensitivity of $SU(1,1)$ interferometer, because $SU(1,2)$
interferometer amplifies the intensity of the input beams twice. As for the
case of one coherent state light input, we analyze the optimal detection
combination when coherent state light is fed into different input ports. It is
shown that the Heisenberg limit can be approached only when the coherent state
light is injected into the third input port.

\begin{acknowledgements}
Y. Wu would like to thank J. Zhang, J. Jing, G. He and W. Zhang for their helpful suggestions. This work is supported by  the National Natural Science Foundation of China (Grants No. 11474095),  and the Fundamental Research Funds for the Central Universities.
\end{acknowledgements}

\bibliographystyle{apsrev4-1}
\bibliography{ref}

\begin{thebibliography}{18}%
\makeatletter
\providecommand \@ifxundefined [1]{%
 \@ifx{#1\undefined}
}%
\providecommand \@ifnum [1]{%
 \ifnum #1\expandafter \@firstoftwo
 \else \expandafter \@secondoftwo
 \fi
}%
\providecommand \@ifx [1]{%
 \ifx #1\expandafter \@firstoftwo
 \else \expandafter \@secondoftwo
 \fi
}%
\providecommand \natexlab [1]{#1}%
\providecommand \enquote  [1]{``#1''}%
\providecommand \bibnamefont  [1]{#1}%
\providecommand \bibfnamefont [1]{#1}%
\providecommand \citenamefont [1]{#1}%
\providecommand \href@noop [0]{\@secondoftwo}%
\providecommand \href [0]{\begingroup \@sanitize@url \@href}%
\providecommand \@href[1]{\@@startlink{#1}\@@href}%
\providecommand \@@href[1]{\endgroup#1\@@endlink}%
\providecommand \@sanitize@url [0]{\catcode `\\12\catcode `\$12\catcode
  `\&12\catcode `\#12\catcode `\^12\catcode `\_12\catcode `\%12\relax}%
\providecommand \@@startlink[1]{}%
\providecommand \@@endlink[0]{}%
\providecommand \url  [0]{\begingroup\@sanitize@url \@url }%
\providecommand \@url [1]{\endgroup\@href {#1}{\urlprefix }}%
\providecommand \urlprefix  [0]{URL }%
\providecommand \Eprint [0]{\href }%
\providecommand \doibase [0]{http://dx.doi.org/}%
\providecommand \selectlanguage [0]{\@gobble}%
\providecommand \bibinfo  [0]{\@secondoftwo}%
\providecommand \bibfield  [0]{\@secondoftwo}%
\providecommand \translation [1]{[#1]}%
\providecommand \BibitemOpen [0]{}%
\providecommand \bibitemStop [0]{}%
\providecommand \bibitemNoStop [0]{.\EOS\space}%
\providecommand \EOS [0]{\spacefactor3000\relax}%
\providecommand \BibitemShut  [1]{\csname bibitem#1\endcsname}%
\let\auto@bib@innerbib\@empty
\bibitem [{\citenamefont {Giovannetti}\ \emph {et~al.}(2004)\citenamefont
  {Giovannetti}, \citenamefont {Lloyd},\ and\ \citenamefont
  {Maccone}}]{giovannetti2004quantum}%
  \BibitemOpen
  \bibfield  {author} {\bibinfo {author} {\bibfnamefont {V.}~\bibnamefont
  {Giovannetti}}, \bibinfo {author} {\bibfnamefont {S.}~\bibnamefont {Lloyd}},
  \ and\ \bibinfo {author} {\bibfnamefont {L.}~\bibnamefont {Maccone}},\
  }\href@noop {} {\bibfield  {journal} {\bibinfo  {journal} {Science}\ }\textbf
  {\bibinfo {volume} {306}},\ \bibinfo {pages} {1330} (\bibinfo {year}
  {2004})}\BibitemShut {NoStop}%
\bibitem [{\citenamefont {Giovannetti}\ \emph {et~al.}(2011)\citenamefont
  {Giovannetti}, \citenamefont {Lloyd},\ and\ \citenamefont
  {Maccone}}]{giovannetti2011advances}%
  \BibitemOpen
  \bibfield  {author} {\bibinfo {author} {\bibfnamefont {V.}~\bibnamefont
  {Giovannetti}}, \bibinfo {author} {\bibfnamefont {S.}~\bibnamefont {Lloyd}},
  \ and\ \bibinfo {author} {\bibfnamefont {L.}~\bibnamefont {Maccone}},\
  }\href@noop {} {\bibfield  {journal} {\bibinfo  {journal} {Nat. Photonics}\
  }\textbf {\bibinfo {volume} {5}},\ \bibinfo {pages} {222} (\bibinfo {year}
  {2011})}\BibitemShut {NoStop}%
\bibitem [{\citenamefont {Caves}(1981)}]{PhysRevD.23.1693}%
  \BibitemOpen
  \bibfield  {author} {\bibinfo {author} {\bibfnamefont {C.~M.}\ \bibnamefont
  {Caves}},\ }\href@noop {} {\bibfield  {journal} {\bibinfo  {journal} {Phys.
  Rev. D}\ }\textbf {\bibinfo {volume} {23}},\ \bibinfo {pages} {1693}
  (\bibinfo {year} {1981})}\BibitemShut {NoStop}%
\bibitem [{\citenamefont {Yurke}\ \emph {et~al.}(1986)\citenamefont {Yurke},
  \citenamefont {McCall},\ and\ \citenamefont {Klauder}}]{PhysRevA.33.4033}%
  \BibitemOpen
  \bibfield  {author} {\bibinfo {author} {\bibfnamefont {B.}~\bibnamefont
  {Yurke}}, \bibinfo {author} {\bibfnamefont {S.~L.}\ \bibnamefont {McCall}}, \
  and\ \bibinfo {author} {\bibfnamefont {J.~R.}\ \bibnamefont {Klauder}},\
  }\href@noop {} {\bibfield  {journal} {\bibinfo  {journal} {Phys. Rev. A}\
  }\textbf {\bibinfo {volume} {33}},\ \bibinfo {pages} {4033} (\bibinfo {year}
  {1986})}\BibitemShut {NoStop}%
\bibitem [{\citenamefont {Plick}\ \emph {et~al.}(2010)\citenamefont {Plick},
  \citenamefont {Dowling},\ and\ \citenamefont {Agarwal}}]{plick2010coherent}%
  \BibitemOpen
  \bibfield  {author} {\bibinfo {author} {\bibfnamefont {W.~N.}\ \bibnamefont
  {Plick}}, \bibinfo {author} {\bibfnamefont {J.~P.}\ \bibnamefont {Dowling}},
  \ and\ \bibinfo {author} {\bibfnamefont {G.~S.}\ \bibnamefont {Agarwal}},\
  }\href@noop {} {\bibfield  {journal} {\bibinfo  {journal} {New J. Phys.}\
  }\textbf {\bibinfo {volume} {12}},\ \bibinfo {pages} {083014} (\bibinfo
  {year} {2010})}\BibitemShut {NoStop}%
\bibitem [{\citenamefont {Li}\ \emph {et~al.}(2014)\citenamefont {Li},
  \citenamefont {Yuan}, \citenamefont {Ou},\ and\ \citenamefont
  {Zhang}}]{li2014phase}%
  \BibitemOpen
  \bibfield  {author} {\bibinfo {author} {\bibfnamefont {D.}~\bibnamefont
  {Li}}, \bibinfo {author} {\bibfnamefont {C.-H.}\ \bibnamefont {Yuan}},
  \bibinfo {author} {\bibfnamefont {Z.}~\bibnamefont {Ou}}, \ and\ \bibinfo
  {author} {\bibfnamefont {W.}~\bibnamefont {Zhang}},\ }\href@noop {}
  {\bibfield  {journal} {\bibinfo  {journal} {New J. Phys.}\ }\textbf {\bibinfo
  {volume} {16}},\ \bibinfo {pages} {073020} (\bibinfo {year}
  {2014})}\BibitemShut {NoStop}%
\bibitem [{\citenamefont {Sanders}\ \emph {et~al.}(1999)\citenamefont
  {Sanders}, \citenamefont {de~Guise}, \citenamefont {Rowe},\ and\
  \citenamefont {Mann}}]{sanders1999vector}%
  \BibitemOpen
  \bibfield  {author} {\bibinfo {author} {\bibfnamefont {B.~C.}\ \bibnamefont
  {Sanders}}, \bibinfo {author} {\bibfnamefont {H.}~\bibnamefont {de~Guise}},
  \bibinfo {author} {\bibfnamefont {D.~J.}\ \bibnamefont {Rowe}}, \ and\
  \bibinfo {author} {\bibfnamefont {A.}~\bibnamefont {Mann}},\ }\href@noop {}
  {\bibfield  {journal} {\bibinfo  {journal} {J. Phys. A: Math Theor}\ }\textbf
  {\bibinfo {volume} {32}},\ \bibinfo {pages} {7791} (\bibinfo {year}
  {1999})}\BibitemShut {NoStop}%
\bibitem [{\citenamefont {Rowe}\ \emph {et~al.}(1999)\citenamefont {Rowe},
  \citenamefont {Sanders},\ and\ \citenamefont
  {de~Guise}}]{rowe1999representations}%
  \BibitemOpen
  \bibfield  {author} {\bibinfo {author} {\bibfnamefont {D.~J.}\ \bibnamefont
  {Rowe}}, \bibinfo {author} {\bibfnamefont {B.~C.}\ \bibnamefont {Sanders}}, \
  and\ \bibinfo {author} {\bibfnamefont {H.}~\bibnamefont {de~Guise}},\
  }\href@noop {} {\bibfield  {journal} {\bibinfo  {journal} {J. Math. Phys.}\
  }\textbf {\bibinfo {volume} {40}},\ \bibinfo {pages} {3604} (\bibinfo {year}
  {1999})}\BibitemShut {NoStop}%
\bibitem [{\citenamefont {Tan}\ \emph {et~al.}(2013)\citenamefont {Tan},
  \citenamefont {Gao}, \citenamefont {de~Guise},\ and\ \citenamefont
  {Sanders}}]{PhysRevLett.110.113603}%
  \BibitemOpen
  \bibfield  {author} {\bibinfo {author} {\bibfnamefont {S.-H.}\ \bibnamefont
  {Tan}}, \bibinfo {author} {\bibfnamefont {Y.~Y.}\ \bibnamefont {Gao}},
  \bibinfo {author} {\bibfnamefont {H.}~\bibnamefont {de~Guise}}, \ and\
  \bibinfo {author} {\bibfnamefont {B.~C.}\ \bibnamefont {Sanders}},\
  }\href@noop {} {\bibfield  {journal} {\bibinfo  {journal} {Phys. Rev. Lett.}\
  }\textbf {\bibinfo {volume} {110}},\ \bibinfo {pages} {113603} (\bibinfo
  {year} {2013})}\BibitemShut {NoStop}%
\bibitem [{\citenamefont {Noh}\ \emph {et~al.}(1992)\citenamefont {Noh},
  \citenamefont {Foug\`eres},\ and\ \citenamefont {Mandel}}]{PhysRevA.46.2840}%
  \BibitemOpen
  \bibfield  {author} {\bibinfo {author} {\bibfnamefont {J.~W.}\ \bibnamefont
  {Noh}}, \bibinfo {author} {\bibfnamefont {A.}~\bibnamefont {Foug\`eres}}, \
  and\ \bibinfo {author} {\bibfnamefont {L.}~\bibnamefont {Mandel}},\
  }\href@noop {} {\bibfield  {journal} {\bibinfo  {journal} {Phys. Rev. A}\
  }\textbf {\bibinfo {volume} {46}},\ \bibinfo {pages} {2840} (\bibinfo {year}
  {1992})}\BibitemShut {NoStop}%
\bibitem [{\citenamefont {D'Ariano}\ and\ \citenamefont
  {Paris}(1997)}]{PhysRevA.55.2267}%
  \BibitemOpen
  \bibfield  {author} {\bibinfo {author} {\bibfnamefont {G.~M.}\ \bibnamefont
  {D'Ariano}}\ and\ \bibinfo {author} {\bibfnamefont {M.~G.~A.}\ \bibnamefont
  {Paris}},\ }\href@noop {} {\bibfield  {journal} {\bibinfo  {journal} {Phys.
  Rev. A}\ }\textbf {\bibinfo {volume} {55}},\ \bibinfo {pages} {2267}
  (\bibinfo {year} {1997})}\BibitemShut {NoStop}%
\bibitem [{\citenamefont {Zukowski}\ \emph {et~al.}(1997)\citenamefont
  {Zukowski}, \citenamefont {Zeilinger},\ and\ \citenamefont
  {Horne}}]{PhysRevA.55.2564}%
  \BibitemOpen
  \bibfield  {author} {\bibinfo {author} {\bibfnamefont {M.}~\bibnamefont
  {Zukowski}}, \bibinfo {author} {\bibfnamefont {A.}~\bibnamefont {Zeilinger}},
  \ and\ \bibinfo {author} {\bibfnamefont {M.~A.}\ \bibnamefont {Horne}},\
  }\href@noop {} {\bibfield  {journal} {\bibinfo  {journal} {Phys. Rev. A}\
  }\textbf {\bibinfo {volume} {55}},\ \bibinfo {pages} {2564} (\bibinfo {year}
  {1997})}\BibitemShut {NoStop}%
\bibitem [{\citenamefont {Ben-Aryeh}(2010)}]{BenAryeh20102863}%
  \BibitemOpen
  \bibfield  {author} {\bibinfo {author} {\bibfnamefont {Y.}~\bibnamefont
  {Ben-Aryeh}},\ }\href@noop {} {\bibfield  {journal} {\bibinfo  {journal}
  {Opt. Commun.}\ }\textbf {\bibinfo {volume} {283}},\ \bibinfo {pages} {2863 }
  (\bibinfo {year} {2010})}\BibitemShut {NoStop}%
\bibitem [{\citenamefont {Spagnolo}\ \emph {et~al.}(2012)\citenamefont
  {Spagnolo}, \citenamefont {Aparo}, \citenamefont {Vitelli}, \citenamefont
  {Crespi}, \citenamefont {Ramponi}, \citenamefont {Osellame}, \citenamefont
  {Mataloni},\ and\ \citenamefont {Sciarrino}}]{spagnolo2012quantum}%
  \BibitemOpen
  \bibfield  {author} {\bibinfo {author} {\bibfnamefont {N.}~\bibnamefont
  {Spagnolo}}, \bibinfo {author} {\bibfnamefont {L.}~\bibnamefont {Aparo}},
  \bibinfo {author} {\bibfnamefont {C.}~\bibnamefont {Vitelli}}, \bibinfo
  {author} {\bibfnamefont {A.}~\bibnamefont {Crespi}}, \bibinfo {author}
  {\bibfnamefont {R.}~\bibnamefont {Ramponi}}, \bibinfo {author} {\bibfnamefont
  {R.}~\bibnamefont {Osellame}}, \bibinfo {author} {\bibfnamefont
  {P.}~\bibnamefont {Mataloni}}, \ and\ \bibinfo {author} {\bibfnamefont
  {F.}~\bibnamefont {Sciarrino}},\ }\href@noop {} {\bibfield  {journal}
  {\bibinfo  {journal} {Sci. Rep.}\ }\textbf {\bibinfo {volume} {2}} (\bibinfo
  {year} {2012})}\BibitemShut {NoStop}%
\bibitem [{\citenamefont {Qin}\ \emph {et~al.}(2014)\citenamefont {Qin},
  \citenamefont {Cao}, \citenamefont {Wang}, \citenamefont {Marino},
  \citenamefont {Zhang},\ and\ \citenamefont {Jing}}]{PhysRevLett.113.023602}%
  \BibitemOpen
  \bibfield  {author} {\bibinfo {author} {\bibfnamefont {Z.}~\bibnamefont
  {Qin}}, \bibinfo {author} {\bibfnamefont {L.}~\bibnamefont {Cao}}, \bibinfo
  {author} {\bibfnamefont {H.}~\bibnamefont {Wang}}, \bibinfo {author}
  {\bibfnamefont {A.~M.}\ \bibnamefont {Marino}}, \bibinfo {author}
  {\bibfnamefont {W.}~\bibnamefont {Zhang}}, \ and\ \bibinfo {author}
  {\bibfnamefont {J.}~\bibnamefont {Jing}},\ }\href@noop {} {\bibfield
  {journal} {\bibinfo  {journal} {Phys. Rev. Lett.}\ }\textbf {\bibinfo
  {volume} {113}},\ \bibinfo {pages} {023602} (\bibinfo {year}
  {2014})}\BibitemShut {NoStop}%
\bibitem [{\citenamefont {Cahn}(2006)}]{cahn2006semi}%
  \BibitemOpen
  \bibfield  {author} {\bibinfo {author} {\bibfnamefont {R.~N.}\ \bibnamefont
  {Cahn}},\ }\href@noop {} {\emph {\bibinfo {title} {Semi-simple Lie algebras
  and their representations}}}\ (\bibinfo  {publisher} {Courier Dover
  Publications},\ \bibinfo {year} {2006})\BibitemShut {NoStop}%
\bibitem [{\citenamefont {Yonezawa}\ \emph {et~al.}(2012)\citenamefont
  {Yonezawa}, \citenamefont {Nakane}, \citenamefont {Wheatley}, \citenamefont
  {Iwasawa}, \citenamefont {Takeda}, \citenamefont {Arao}, \citenamefont
  {Ohki}, \citenamefont {Tsumura}, \citenamefont {Berry}, \citenamefont
  {Ralph}, \citenamefont {Wiseman}, \citenamefont {Huntington},\ and\
  \citenamefont {Furusawa}}]{yonezawa2012quantum}%
  \BibitemOpen
  \bibfield  {author} {\bibinfo {author} {\bibfnamefont {H.}~\bibnamefont
  {Yonezawa}}, \bibinfo {author} {\bibfnamefont {D.}~\bibnamefont {Nakane}},
  \bibinfo {author} {\bibfnamefont {T.~A.}\ \bibnamefont {Wheatley}}, \bibinfo
  {author} {\bibfnamefont {K.}~\bibnamefont {Iwasawa}}, \bibinfo {author}
  {\bibfnamefont {S.}~\bibnamefont {Takeda}}, \bibinfo {author} {\bibfnamefont
  {H.}~\bibnamefont {Arao}}, \bibinfo {author} {\bibfnamefont {K.}~\bibnamefont
  {Ohki}}, \bibinfo {author} {\bibfnamefont {K.}~\bibnamefont {Tsumura}},
  \bibinfo {author} {\bibfnamefont {D.~W.}\ \bibnamefont {Berry}}, \bibinfo
  {author} {\bibfnamefont {T.~C.}\ \bibnamefont {Ralph}}, \bibinfo {author}
  {\bibfnamefont {H.~M.}\ \bibnamefont {Wiseman}}, \bibinfo {author}
  {\bibfnamefont {E.~H.}\ \bibnamefont {Huntington}}, \ and\ \bibinfo {author}
  {\bibfnamefont {A.}~\bibnamefont {Furusawa}},\ }\href@noop {} {\bibfield
  {journal} {\bibinfo  {journal} {Science}\ }\textbf {\bibinfo {volume}
  {337}},\ \bibinfo {pages} {1514} (\bibinfo {year} {2012})}\BibitemShut
  {NoStop}%
\bibitem [{\citenamefont {Marino}\ \emph {et~al.}(2012)\citenamefont {Marino},
  \citenamefont {Corzo~Trejo},\ and\ \citenamefont
  {Lett}}]{PhysRevA.86.023844}%
  \BibitemOpen
  \bibfield  {author} {\bibinfo {author} {\bibfnamefont {A.~M.}\ \bibnamefont
  {Marino}}, \bibinfo {author} {\bibfnamefont {N.~V.}\ \bibnamefont
  {Corzo~Trejo}}, \ and\ \bibinfo {author} {\bibfnamefont {P.~D.}\ \bibnamefont
  {Lett}},\ }\href@noop {} {\bibfield  {journal} {\bibinfo  {journal} {Phys.
  Rev. A}\ }\textbf {\bibinfo {volume} {86}},\ \bibinfo {pages} {023844}
  (\bibinfo {year} {2012})}\BibitemShut {NoStop}%
\end{thebibliography}%

\end{document}